\documentclass[a4paper,11pt]{article}
\pdfoutput=1 

\usepackage{jheppub} 

\usepackage[T1]{fontenc} 

\title{\boldmath Sub-sub-leading soft-graviton theorem\\ in arbitrary dimension}

\author[a]{Michael Zlotnikov}

\affiliation[a]{Brown University\\Department of Physics\\182 Hope St, Providence, RI, 02912}

\emailAdd{michael\_zlotnikov@brown.edu}

\abstract{\\The CHY formula which describes an $n$-point tree level scattering amplitude for scalars, gluons or gravitons in arbitrary dimension \cite{chyexp}, is used to prove the sub-sub-leading soft-graviton theorem recently proposed by Cachazo and Strominger \cite{cachstro}.}

\begin{document} 
\maketitle
\flushbottom

\section{Introduction}
A lot of work has been done on soft theorems in the past, based on local on-shell gauge invariance \cite{LowFourPt1,LowFourPt2,LowFourPt3,LowTheorem,Weinberg1,weinberg,OtherSoftPhotons1,OtherSoftPhotons2,OtherSoftPhotons3}. The leading soft-graviton behavior was found by Weinberg in 1965 \cite{weinberg}, and the sub-leading soft-graviton theorem was first investigated by Gross and Jackiw in 1968 \cite{Gross:1968in}. Recently, active interest in soft theorems has been reawakened in \cite{strominger1,strominger2,strominger3,cachstro}, as Strominger and collaborators discovered that soft-graviton behavior can be extracted from extended BMS symmetry \cite{BMS1,BMS2,ExtendedBMS1,ExtendedBMS2,ExtendedBMS3}. For four dimensions, Cachazo and Strominger provided a proof for the universality of tree level sub-leading and sub-sub-leading corrections \cite{cachstro} to Weinbergs soft-graviton factor \cite{weinberg}, making use of spinor helicity formalism and BCFW recursion \cite{Britto:2004ap,bcfwa}. The soft-graviton factor refers to the factorization property of an $(n+1)$-point tree level scattering amplitude when the momentum of one external particle, conventionally the (n+1)\textsuperscript{th} particle, is going to zero\footnote{Substitute $k_{n+1}\to \varepsilon k_{n+1}$ and expand around $\varepsilon=0$.}
\begin{align}
 \mathcal{M}_{n+1}(k_1,k_2,...,\varepsilon k_{n+1})=\left(\frac{1}{\varepsilon}S^{(0)}+S^{(1)}+\varepsilon S^{(2)}+\mathcal{O}(\varepsilon^2)\right)\mathcal{M}_n(k_1,k_2,...,k_n).
\end{align}
In case of gravity, these soft factors read \cite{cachstro}
\begin{subequations}
\label{eq:factors}
\begin{align}
\label{eq:lead}
S^{(0)}=&\sum_{a=1}^n\frac{{\epsilon_{n+1}}_{\mu\nu} k^\mu_a k^\nu_a}{k_{n+1}\cdot k_a}\\
\label{eq:sublead}
S^{(1)}=&\sum_{a=1}^n\frac{{\epsilon_{n+1}}_{\mu\nu}  k^\mu_a ({k_{n+1}}_\lambda J^{\lambda \nu}_a)}{k_{n+1}\cdot k_a}\\
\label{eq:subsublead}
S^{(2)}=&\frac{1}{2}\sum_{a=1}^n\frac{{\epsilon_{n+1}}_{\mu\nu}  ({k_{n+1}}_\rho J^{\rho \mu}_a) ({k_{n+1}}_\lambda J^{\lambda \nu}_a)}{k_{n+1}\cdot k_a},
\end{align}
\end{subequations}
where ${\epsilon_{n+1}}_{\mu\nu}$ is the polarization tensor of the $(n+1)$\textsuperscript{th} particle, $k^\mu_a$ are momenta and $J_a^{\mu\nu}$ are angular momentum operators. Subsequently, these soft-graviton theorems are being investigated with the restriction to four dimensions lifted. In arbitrary number of dimensions, the leading factor (\ref{eq:lead}) was addressed in \cite{chyexp} and the sub-leading factor (\ref{eq:sublead}) was explicitly confirmed in \cite{nima,schwab}. Considering Poincar\'e and gauge invariance in arbitrary number of dimensions, as well as expected formal structure, Broedel, de Leeuw, Plefka and Rosso fixed the orbital part of the sub-leading and sub-sub-leading factors completely and constrained their polarization parts up to one numerical constant for every order of expansion and each hard leg  \cite{Broedel:2014fsa}, in agreement with (\ref{eq:factors}). Following Low's example \cite{LowTheorem}, Bern, Davies, Di Vecchia and Nohle used on-shell gauge invariance to fully determine and confirm the first two sub-leading soft-graviton behaviors in $D$ dimensions \cite{Bern:2014vva}.

Further work on soft factors in general was, for instance, done for Yang-Mills amplitudes in \cite{Casali:2014xpa},\cite{Larkoski:2014hta}. Several advances in gauge and gravity theories at loop level appeared in \cite{Bern:2014oka,He:2014bga}.\footnote{Nontrivial corrections are expected at loop level.} Cachazo and Yuan proposed a modification of the usual soft limit procedure to cope with corrections appearing at loop level  \cite{Cachazo:2014dia}. For a comment on this procedure, see \cite{Bern:2014vva}. Sub-leading soft theorems in gauge and gravity theory were confirmed from a diagrammatic approach in \cite{White:2011yy,White:2014qia}. Soft theorem in QED was revisited in \cite{He:2014cra,Lysov:2014csa}. Stringy soft theorems appeared in \cite{Geyer:2014lca,Schwab:2014fia}, and a more general investigation of soft theorems in a broader set of theories was conducted in \cite{Bianchi:2014gla}.

In this note we will contribute an additional proof of the validity of the sub-sub-leading tree level soft factor (\ref{eq:subsublead}) by explicit computation in arbitrary dimension, making use of the CHY formula \cite{chyexp}. This note is structured as follows. Section \ref{sec:chy} recalls the CHY formula. In section \ref{sec:hpe} we outline the computational steps for the higher point expansion in the soft limit. Section \ref{sec:lpc} contains the computation of lower point construction and comparison of the two results. Appendices \ref{app:d2d0}, \ref{app:d1d1} and \ref{app:d0d2} contain all terms resulting from higher point expansion, which are compared with and are found to be equal to the result of lower point construction.

\section{The CHY formula}
\label{sec:chy}
In order to explicitly prove the sub-sub-leading factor in the soft-graviton expansion, we will make use of the CHY formula for tree level gravity scattering amplitudes with $(n+1)$ external legs, which is valid in any number of dimensions \cite{chyexp}
\begin{align}
\mathcal{M}_{n+1}=\int\left[\prod_{{c=1}\atop{c\neq p,q,r}}^{n+1}d\sigma_c\right]\frac{4(\sigma_{pq}\sigma_{qr}\sigma_{rp})(\sigma_{ij}\sigma_{jk}\sigma_{ki})}{(\sigma_{mw})^2}\left[\prod_{{a=1}\atop{a\neq i,j,k}}^{n+1}{\delta\left(\sum_{{b=1}\atop{b\neq a}}^{n+1}\frac{k_a\cdot k_b}{\sigma_{ab}}\right)}\right] \det\left(\Psi^{m,w}_{m,w}\right).\notag
\end{align}
Here we use the abbreviation $\sigma_{ij}\equiv(\sigma_i-\sigma_j)$. Upper indices on the matrix $\Psi$ denote removed columns and lower indices denote removed rows. Values of indices $p,q,r,i,j,k,m$ and $w$ can be chosen arbitrarily without changing the result. The $2(n+1)$ dimensional matrix $\Psi$ is given by
\begin{align}
\Psi=\left({{A}\atop{C}}~{{-C^T}\atop{B}}\right),\notag
\end{align}
where the $(n+1)$ dimensional sub-matrices are given by
\begin{align}
A_{ab}=\left\{{{\frac{k_a\cdot k_b}{\sigma_{ab}}}\atop{0}}{{,~a\neq b}\atop{,~a=b}}\right.~~~,~~~B_{ab}=\left\{{{\frac{\epsilon_a\cdot \epsilon_b}{\sigma_{ab}}}\atop{0}}{{,~a\neq b}\atop{,~a=b}}\right.~~~,~~~C_{ab}=\left\{{{\frac{\epsilon_a\cdot k_b}{\sigma_{ab}}}\atop{-\sum_{{c=1}\atop{c\neq a}}^{n+1}\frac{\epsilon_a\cdot k_c}{\sigma_{ac}}}}{{,~a\neq b}\atop{,~a=b}}\right..\notag
\end{align}
Here $k_a^\mu$ is the momentum of the $a$\textsuperscript{th} particle, and $\epsilon_a^\mu$ is part of its polarization tensor. The values for all $\sigma_i$ in the integration are fixed by the product of delta functions which enforce the scattering equations.
The momentum of the $(n+1)$\textsuperscript{th} leg will be sent to zero in the soft-graviton expansion.

\section{Higher point expansion}
\label{sec:hpe}
In the higher point expansion we start with the momentum conservation stripped tree level amplitude for $n+1$ external gravitons, substitute $k_{n+1}^\mu\to\varepsilon k_{n+1}^\mu$ and expand around $\varepsilon=0$. In the sub-sub-leading case we are interested in the order $\mathcal{O}(\varepsilon^1)$ terms of this expansion. Subsequently, we integrate out the $\sigma_{n+1}$ dependance to obtain the result which we expect to recover from lower point construction by acting with the corresponding soft factor on an amplitude with one fewer external leg in section \ref{sec:lpc}. All solutions for $\sigma_{n+1}$ are fixed by the scattering equations. However, since we are dealing with tree level amplitudes, the functional dependance does not feature any branch cuts such that we will be able to deform the integration contour and pick up a different set of residues in $\sigma_{n+1}$ as in \cite{nima,schwab} in order to obtain the same result, effectively avoiding having to solve the scattering equations.

For convenience we set $i=1$, $j=n$, $m=2$ and $w=3$ so that the momentum conservation stripped tree level amplitude for $n+1$ external gravitons is given by
\begin{align*}
\mathcal{M}_{n+1}=\int\left[\prod_{{c=1}\atop{c\neq p,q,r}}^{n+1}d\sigma_c\right]\frac{4(\sigma_{pq}\sigma_{qr}\sigma_{rp})(\sigma_{1n}\sigma_{nk}\sigma_{k1})}{(\sigma_{2,3})^2}\left[\prod_{{a=2}\atop{a\neq k,n}}^{n+1}{\delta\left(\sum_{{b=1}\atop{b\neq a}}^{n+1}\frac{k_a\cdot k_b}{\sigma_{ab}}\right)}\right] \det\left(\Psi^{2,3}_{2,3}\right).
\end{align*}
Since tree level amplitudes do not feature branch cuts, a delta distribution can be mapped to a single pole term
\begin{align*}
  \delta\left(\sum_{{b=1}\atop{b\neq a}}^{n+1}\frac{k_a\cdot k_b}{\sigma_{ab}}\right) \to \frac{1}{\sum_{{b=1}\atop{b\neq a}}^{n+1}\frac{k_a\cdot k_b}{\sigma_{ab}}}
\end{align*}
 while the integration contour is deformed to pick up the residue associated with this pole as in \cite{nima,schwab}. This naturally yields the same result for the amplitude. Therefore, we can substitute one delta function that has index $a=(n+1)$ by a simple pole, take $k_{n+1}\to \varepsilon k_{n+1}$ and expand around $\varepsilon=0$ as follows:
\begin{align}
\label{eq:delta}
	&\prod_{{a=2}\atop{a\neq k,n}}^{n+1}{
		\delta\left(\sum_{{b=1}\atop{b\neq a}}^{n+1}\frac{k_a\cdot k_b}{\sigma_{ab}}\right)
	} 
	= \\
	&=\,
	\frac{1}{\sum_{{c=1}}^{n}\frac{k_{n+1}\cdot k_c}{\sigma_{n+1,c}}} 
	\left[
		\frac{1}{\varepsilon}\prod_{{a=2}\atop{a\neq k}}^{n-1}{
			\delta\left(\sum_{{b=1}\atop{b\neq a}}^{n}\frac{k_a\cdot k_b}{\sigma_{ab}}\right)
		} 
	  +\sum_{{r=2}\atop{r\neq k}}^{n-1}\frac{k_{n+1}\cdot k_r}{\sigma_{n+1,r}} 
		\delta^{(1)}\left(
			\sum_{{q=1}\atop{q\neq r}}^{n}\frac{k_r\cdot k_q}{\sigma_{r q}}
		\right)
		\prod_{{a=2}\atop{a\neq k,r}}^{n-1}{
		  \delta\left(\sum_{{b=1}\atop{b\neq a}}^{n}\frac{k_a\cdot k_b}{\sigma_{ab}}\right)
	  } \right. \notag\\
	&~~~
		+\frac{\varepsilon}{2}\sum_{{r=2}\atop{r\neq k}}^{n-1}\frac{k_{n+1}\cdot k_r}{\sigma_{n+1,r}} 
		\delta^{(1)}\left(
			\sum_{{q=1}\atop{q\neq r}}^{n}\frac{k_r\cdot k_q}{\sigma_{r q}}
		\right)
		\sum_{{s=2}\atop{s\neq k,r}}^{n-1}\frac{k_{n+1}\cdot k_s}{\sigma_{n+1,s}} 
		\delta^{(1)}\left(
			\sum_{{t=1}\atop{t\neq s}}^{n}\frac{k_s\cdot k_t}{\sigma_{s t}}
		\right)
		\prod_{{a=2}\atop{a\neq k,r,s}}^{n-1}{
		  \delta\left(\sum_{{b=1}\atop{b\neq a}}^{n}\frac{k_a\cdot k_b}{\sigma_{ab}}\right)
	  } \notag\\
	&~~~
	  \left. +\frac{\varepsilon}{2}\sum_{{r=2}\atop{r\neq k}}^{n-1}\left(\frac{k_{n+1}\cdot k_r}{\sigma_{n+1,r}}\right)^2
		\delta^{(2)}\left(
			\sum_{{q=1}\atop{q\neq r}}^{n}\frac{k_r\cdot k_q}{\sigma_{r q}}
		\right)
		\prod_{{a=2}\atop{a\neq k,r}}^{n-1}{
		  \delta\left(\sum_{{b=1}\atop{b\neq a}}^{n}\frac{k_a\cdot k_b}{\sigma_{ab}}\right)
	  } + \mathcal{O}(\varepsilon^2) 
	\right] \notag\\
	&\equiv\,\frac{1}{\varepsilon}\delta^0+\delta^1+\varepsilon\delta^2 +\mathcal{O}(\varepsilon^2).\notag
\end{align}
Here we have introduced abbreviations $\delta^i$ to denote the expansion coefficients of order $\varepsilon^{i-1}$. Similarly, we can expand the determinant $\det (\Psi^{2,3}_{2,3} )$ to make its $\varepsilon$ dependance explicit. For that end we employ the usual recursive formula\footnote{In this case we are dealing with a $2(n+1)\times 2(n+1)$ matrix.}
\begin{align}
\label{eq:detexp}
\det(A)=\sum_{i=1}^{2(n+1)}(-1)^{i+k}a_{ki}\det(A^i_k),
\end{align}
where $a_{ki}$ are elements of matrix $A$ and the choice of row $k$ is arbitrary.\footnote{Naturally, an analogous expansion can also be done along a column instead of a row.} If certain rows and columns are initially missing from the matrix $A$ such that it is less than $2(n+1)\times 2(n+1)$ dimensional before the expansion (\ref{eq:detexp}) is applied, those corresponding values of missing rows and columns have to be skipped in the summation over the expansion index $i$. Additionally, a jump by $\pm1$ has to be introduced in the exponent of $(-1)^{i+k}$ whenever such a missing row or column is crossed. This will be accomplished with help of the Heaviside step function $\theta(a,b)\equiv\theta(a-b)$. More explicitly, when an additional row $i$ (or column $k$) is removed from matrix $A$, one step function has to be introduced for each of the rows $u$ (or columns $v$)  that were already missing, so that we add\footnote{Note that the newly removed index $i$ (or $k$) is in the first argument of each respective step function and is attached to the determinant at the far right. This introduces a natural initial index-ordering and will be relevant in the following.} $\sum_u \theta(i,u)$ (or $\sum_v \theta(k,v)$) to the exponent of $(-1)^{i+k}$. This ensures that each summand in the expansion (\ref{eq:detexp}) appears with the correct sign.\\
As in \cite{nima}, we make use of the gauge condition $(k_{n+1}\cdot\epsilon_i)=0$ for all $i$ to conveniently reduce the number of appearing terms. We realize that with this all the $\varepsilon$ dependance is located along the $(n+1)$\textsuperscript{th} row and column of $\Psi^{2,3}_{2,3}$. Therefore, we apply the expansion to the $(n+1)$\textsuperscript{th} row and column in succession:
\begin{align}
\label{eq:exp1}
\det \left(\Psi^{2,3}_{2,3}\right)=&\left(\sum_{c=1}^{n}\frac{\epsilon_{n+1}\cdot k_c}{\sigma_{n+1,c}}\right)^2\det \left(\Psi^{2,3,n+1,2(n+1)}_{2,3,n+1,2(n+1)}\right) \\
&+2\varepsilon\sum_{{i=1}\atop{i\neq2,3}}^{n}\sum_{c=1}^{n}(-1)^i\frac{\epsilon_{n+1}\cdot k_c}{\sigma_{n+1,c}}\frac{k_{n+1}\cdot k_i}{\sigma_{n+1,i}}\det\left(\Psi^{2,3,n+1,i}_{2,3,n+1,2(n+1)}\right)\notag\\
&+\varepsilon^2\sum_{{i=1}\atop{i\neq2,3}}^{n}\sum_{{j=1}\atop{j\neq2,3}}^{n}(-1)^{i+j}\frac{k_{n+1}\cdot k_i}{\sigma_{n+1,i}}\frac{k_j\cdot k_{n+1}}{\sigma_{n+1,j}}\det\left(\Psi^{2,3,n+1,i}_{2,3,n+1,j}\right).\notag
\end{align}
Here and in later equations the Heaviside step functions involving arguments $2$, $3$, $n+1$ and $2(n+1)$ are suppressed. However, to keep track of the signs we should agree to always order the argument of each step function according to the order in which removed rows or columns appear in the determinant. In particular, 
\begin{align}
\label{eq:theta}
(-1)^{...+\theta(a,b)+...+\theta(c,d)+...}\det(\Psi^{...,b,a,...}_{...,d,c,...})&=-(-1)^{...+\theta(b,a)+...+\theta(c,d)+...}\det(\Psi^{...,a,b,...}_{...,d,c,...})\\
&=(-1)^{...+\theta(b,a)+...+\theta(d,c)+...}\det(\Psi^{...,a,b,...}_{...,c,d,...}).\notag
\end{align}
In cases where more than two rows (columns) are removed from $\Psi$, there will be one step function for each way an unordered pair of removed rows (columns) can be selected. Therefore, with our agreement (\ref{eq:theta}) we can think of the step functions as being attached to the determinant, facilitating the property of making the exchange of two neighboring indices of removed rows (or columns) antisymmetric. Furthermore, this ensures that the order of arguments of all step functions is in one to one correspondence to the order of removed row (column) indices in the determinant, allowing us to ignore the step functions and concentrate on comparing determinants. This convenient property yields a slight simplification to the algebraic steps that later will be required in order to show the equality of the higher point expansion and lower point construction results.\footnote{Some of the appearing step functions can never yield a change of sign and it might be tempting to evaluate them right away and get rid of them. However, this would break the agreement (\ref{eq:theta}) and the convenient general antisymmetry property of the determinant under exchange of two neighboring removed row (column) indices, thus making a more tedious case by case distinction for index-ordering necessary.}\\
Note that the order in which the indices of removed rows and columns appear in the determinants in (\ref{eq:exp1}) is different from the straightforward order which emerges from the expansion. We reordered these indices according to (\ref{eq:theta}) to ensure proper sign in comparison to the terms of lower point construction computed in the next section. For later convenience we define the abbreviation:
\begin{align}
\label{eq:detabbr}
\det \left(\Psi'\right)\equiv\det \left(\Psi^{2,3,n+1,2(n+1)}_{2,3,n+1,2(n+1)}\right).
\end{align}
We wish to make the entire $\sigma_{n+1}$ dependance explicit to be able to integrate it out. Only $(n+1)$\textsuperscript{th} and $2(n+1)$\textsuperscript{th} rows and columns in the matrix $\Psi$ depend on $\sigma_{n+1}$. Therefore, we expand $\det(\Psi^{2,3,n+1,i}_{2,3,n+1,2(n+1)})$ along the $2(n+1)$\textsuperscript{th} column, as well as $\det(\Psi^{2,3,n+1,i}_{2,3,n+1,j})$ along the $2(n+1)$\textsuperscript{th} row and column in succession. Again, here and in all further steps we make use of the gauge condition $(k_{n+1}\cdot\epsilon_i)=0$ for all $i$, such that:
\begin{align}
\label{eq:detfullexp}
\det \left(\Psi^{2,3}_{2,3}\right)=&\left(\sum_{c=1}^{n}\frac{\epsilon_{n+1}\cdot k_c}{\sigma_{n+1,c}}\right)^2\det \left(\Psi'\right) +2\varepsilon\sum_{{i=1}\atop{i\neq2,3}}^{n}\sum_{c=1}^{n}(-1)^i\frac{\epsilon_{n+1}\cdot k_c}{\sigma_{n+1,c}}\frac{k_{n+1}\cdot k_i}{\sigma_{n+1,i}}\times\\
&~~~\times\left(\sum_{{j=1}\atop{j\neq2,3}}^{n}(-1)^j\frac{\epsilon_{n+1}\cdot k_j}{\sigma_{n+1,j}}\det\left(\Psi'^i_j\right)-\sum_{j=1}^{n}(-1)^{j+n+1}\frac{\epsilon_{n+1}\cdot\epsilon_j}{\sigma_{n+1,j}}\det\left(\Psi'^i_{j+n+1}\right)\right)\notag\\
&+\varepsilon^2\sum_{{i=1}\atop{i\neq2,3}}^{n}\sum_{{j=1}\atop{j\neq2,3}}^{n}(-1)^{i+j}\frac{k_{n+1}\cdot k_i}{\sigma_{n+1,i}}\frac{k_j\cdot k_{n+1}}{\sigma_{n+1,j}}\times\notag\\
&~~~\times\left(\sum_{{u=1}\atop{u\neq2,3,i}}^{n}\sum_{{p=1}\atop{p\neq2,3,j}}^{n}(-1)^{u+p+\theta(u,i)+\theta(p,j)}\frac{\epsilon_{n+1}\cdot k_u}{\sigma_{n+1,u}}\frac{\epsilon_{n+1}\cdot k_p}{\sigma_{n+1,p}}\det\left(\Psi'^{i,u}_{j,p}\right)\right.\notag\\
&~~~~~~~-2\sum_{{u=1}\atop{u\neq2,3,i}}^{n}\sum_{{p=1}}^{n}(-1)^{u+p+n+1+\theta(u,i)+\theta(p+n+1,j)}\frac{\epsilon_{n+1}\cdot k_u}{\sigma_{n+1,u}}\frac{\epsilon_{n+1}\cdot \epsilon_p}{\sigma_{n+1,p}}\det\left(\Psi'^{i,u}_{j,p+n+1}\right)\notag\\
&~~~~~~~\left.+\sum_{{u=1}}^{n}\sum_{{p=1}}^{n}(-1)^{u+p+\theta(u+n+1,i)+\theta(p+n+1,j)}\frac{\epsilon_{n+1}\cdot \epsilon_u}{\sigma_{n+1,u}}\frac{\epsilon_{n+1}\cdot \epsilon_p}{\sigma_{n+1,p}}\det\left(\Psi'^{i,u+n+1}_{j,p+n+1}\right)\right)\notag\\
\equiv&~{\det}_0+\varepsilon\,{\det}_1+\varepsilon^2\,{\det}_2.\notag
\end{align}
Here, again we defined abbreviations ${\det}_i$ to denote the coefficients of $\varepsilon^i$. The ordering of the indices of removed rows and columns in the determinants was again done in accordance with (\ref{eq:theta}) to ensure proper signs. In the sub-sub-leading case at hand only terms of overall order $\mathcal{O}(\varepsilon^1)$ are of interest. Therefore, we restrict our attention to:
\begin{align}
\prod_{{a=2}\atop{a\neq k,n}}^{n+1}{
		\delta\left(\sum_{{b=1}\atop{b\neq a}}^{n+1}\frac{k_a\cdot k_b}{\sigma_{ab}}\right)
	} \det \left(\Psi^{2,3}_{2,3}\right)=&\left(\frac{1}{\varepsilon}\delta^0+\delta^1+\varepsilon\delta^2 + \mathcal{O}(\varepsilon^2)\right)\left({\det}_0+\varepsilon\,{\det}_1+\varepsilon^2\,{\det}_2\right) \notag\\
	\label{eq:order}
	=&~\varepsilon\left(\delta^2{\det}_0+\delta^1{\det}_1+\delta^0{\det}_2\right)+...~,
\end{align}
where $...$ denotes other terms of different order in $\varepsilon$. In fact, the terms given explicitly in (\ref{eq:order}) are the only terms of order $\mathcal{O}(\varepsilon^1)$ in the amplitude which depend on $\sigma_{n+1}$. Other multiplicative terms and integrals are merely spectators and can be suppressed when we integrate out $\sigma_{n+1}$ and compare the result to the lower point construction.\\
As in \cite{nima,schwab}, it is trivial to see that there is no pole and therefore no residue at infinity in $\sigma_{n+1}$. Therefore, the integration contour can be reversed to pick up the residues at $\sigma_{n+1}=\sigma_i$ for all $i\neq n+1$ instead. Poles of higher order will occur in the computation, so that we will use Cauchy's integral formula to obtain the respective residues:
\begin{align}
\label{eq:cauchy}
\text{Res}\left(\frac{f(z)}{(z-z_0)^n}~,~z=z_0\right)=\frac{1}{(n-1)!}f^{(n-1)}(z_0),
\end{align}
where $f^{(n-1)}(z)$ is the $(n-1)$\textsuperscript{th} derivative of $f(z)$.\\
The technical steps necessary to obtain the residues from all the terms of order $\mathcal{O}(\varepsilon^1)$ appearing in (\ref{eq:order}) are identical. Let us illustrate the procedure on one expression from $\delta^2{\det}_0$:
\begin{align}
\label{eq:hpex}
\frac{1}{2}\frac{\left(\sum_{b=1}^{n}\frac{\epsilon_{n+1}\cdot k_b}{\sigma_{n+1,b}}\right)^2}{\sum_{c=1}^{n}\frac{k_{n+1}\cdot k_c}{\sigma_{n+1,c}}}\sum_{{r=2}\atop{r\neq k}}^{n-1}\left(\frac{k_{n+1}\cdot k_r}{\sigma_{n+1,r}}\right)^2\delta^{(2)}\left(\sum_{{t=1}\atop{t\neq r}}^{n}\frac{k_{r}\cdot k_t}{\sigma_{r,t}}\right)
\prod_{{a=2}\atop{a\neq k,r}}^{n-1}{
	\delta\left(\sum_{{p=1}\atop{p\neq a}}^{n}\frac{k_{a}\cdot k_p}{\sigma_{a,p}}\right)
}\det\left(\Psi'\right).
\end{align}
First, we suppress the product of delta functions and the determinant since they are just spectators independent of $\sigma_{n+1}$, and we abbreviate $\delta^{(2)}_r\equiv\delta^{(2)}(\sum_{{t=1}\atop{t\neq r}}^{n}\frac{k_{r}\cdot k_t}{\sigma_{r,t}})$ for convenience. To investigate the residues at $\sigma_{n+1}=\sigma_i$ for all $i\neq n+1$, it is natural to distinguish between two cases of $\sigma_{n+1}=\sigma_q$ where $\sigma_q\in\{\sigma_1,\sigma_k,\sigma_n\}$ and $\sigma_{q}\notin\{\sigma_1,\sigma_k,\sigma_n\}$. In the first case, where $\sigma_q\in\{\sigma_1,\sigma_k,\sigma_n\}$ we find only first order poles in (\ref{eq:hpex}):
\begin{align*}
\frac{1}{\sigma_{n+1,q}}\left(\frac{1}{2}\frac{\left(\epsilon_{n+1}\cdot k_q+\sigma_{n+1,q}\sum_{{b=1}\atop{b\neq q}}^{n}\frac{\epsilon_{n+1}\cdot k_b}{\sigma_{n+1,b}}\right)^2}{k_{n+1}\cdot k_q+\sigma_{n+1,q}\sum_{{c=1}\atop{c\neq q}}^{n}\frac{k_{n+1}\cdot k_c}{\sigma_{n+1,c}}}\sum_{{r=2}\atop{r\neq k}}^{n-1}\left(\frac{k_{n+1}\cdot k_r}{\sigma_{n+1,r}}\right)^2\delta^{(2)}_r\right),
\end{align*}
so that the sum of corresponding residues is trivially given by using the Cauchy integral formula (\ref{eq:cauchy}) with $n=1$ and summing over $q$:
\begin{align*}
\frac{1}{2}\sum_{q=1,k,n}\frac{\left(\epsilon_{n+1}\cdot k_q\right)^2}{k_{n+1}\cdot k_q}\sum_{{r=2}\atop{r\neq k}}^{n-1}\left(\frac{k_{n+1}\cdot k_r}{\sigma_{q,r}}\right)^2\delta^{(2)}_r.
\end{align*}
In the second case $\sigma_q\notin\{\sigma_1,\sigma_k,\sigma_n\}$ we find first, second and third order poles in (\ref{eq:hpex}):
\begin{align*}
&\frac{1}{\sigma_{n+1,q}}\left(
	\frac{1}{2}\frac{\left(\sum_{{b=1}\atop{b\neq q}}^{n}\frac{\epsilon_{n+1}\cdot k_c}{\sigma_{n+1,c}}\right)^2(k_q\cdot k_{n+1})^2\delta^{(2)}_q}{k_{n+1}\cdot k_q+\sigma_{n+1,q}\sum_{{c=1}\atop{c\neq q}}^n\frac{k_{n+1}\cdot k_c}{\sigma_{n+1,c}}}+
	\frac{1}{2}\frac{(\epsilon_{n+1}\cdot k_q)^2\sum_{{r=2}\atop{r\neq k,q}}^{n-1}\left(\frac{k_r\cdot k_{n+1}}{\sigma_{n+1,r}}\right)^2\delta^{(2)}_r}{k_{n+1}\cdot k_q+\sigma_{n+1,q}\sum_{{c=1}\atop{c\neq q}}^n\frac{k_{n+1}\cdot k_c}{\sigma_{n+1,c}}}
\right) \\
&+\frac{1}{(\sigma_{n+1,q})^2}\left(\frac{(\epsilon_{n+1}\cdot k_q)(k_q\cdot k_{n+1})^2\delta^{(2)}_q \sum_{{b=1}\atop{b\neq q}}^n\frac{\epsilon_{n+1}\cdot k_b}{\sigma_{n+1,b}}}{k_{n+1}\cdot k_q + \sigma_{n+1,q}\sum_{{c=1}\atop{c\neq q}}^n\frac{k_{n+1}\cdot k_c}{\sigma_{n+1,c}}}\right) \\
&+\frac{1}{(\sigma_{n+1,q})^3}\left(\frac{1}{2}\frac{(\epsilon_{n+1}\cdot k_q)^2(k_q\cdot k_{n+1})^2\delta^{(2)}_q}{k_{n+1}\cdot k_q+\sigma_{n+1,q}\sum_{{c=1}\atop{c\neq q}}^n\frac{k_{n+1}\cdot k_c}{\sigma_{n+1,c}}}\right).
\end{align*}
The sum of all simple pole residues again is trivially obtained by using the Cauchy integral formula (\ref{eq:cauchy}) with $n=1$ and summing over $q$:
\begin{align*}
	\frac{1}{2}\sum_{{q=2}\atop{q\neq k}}^{n-1}\left((k_q\cdot k_{n+1})\left(\sum_{{b=1}\atop{b\neq q}}^{n}\frac{\epsilon_{n+1}\cdot k_c}{\sigma_{q,c}}\right)^2\delta^{(2)}_q+\frac{(\epsilon_{n+1}\cdot k_q)^2}{k_{n+1}\cdot k_q}\sum_{{r=2}\atop{r\neq k,q}}^{n-1}\left(\frac{k_r\cdot k_{n+1}}{\sigma_{q,r}}\right)^2\delta^{(2)}_r\right).
\end{align*}
To obtain the sum over second order pole residues we make use of the Cauchy integral formula (\ref{eq:cauchy}) with $n=2$. This yields:
\begin{align*}
-\sum_{{q=2}\atop{q\neq k}}^{n-1}(\epsilon_{n+1}\cdot k_q)\left((k_q\cdot k_{n+1})\sum_{{b=1}\atop{b\neq q}}^n\frac{\epsilon_{n+1}\cdot k_b}{(\sigma_{q,b})^2}\delta^{(2)}_q+\sum_{{b=1}\atop{b\neq q}}^n\frac{\epsilon_{n+1}\cdot k_b}{\sigma_{q,b}}\sum_{{c=1}\atop{c\neq q}}^n\frac{k_{n+1}\cdot k_c}{\sigma_{q,c}}\delta^{(2)}_q\right).
\end{align*}
Finally, to obtain the sum over the third order pole residues we use the Cauchy integral formula (\ref{eq:cauchy}) with $n=3$:
\begin{align*}
\frac{1}{2}\sum_{{q=2}\atop{q\neq k}}^{n-1}(\epsilon_{n+1}\cdot k_q)^2\left(\frac{1}{k_{n+1}\cdot k_q}\left(\sum_{{b=1}\atop{b\neq q}}^n\frac{k_{n+1}\cdot k_b}{\sigma_{q,b}}\right)^2\delta^{(2)}_q+\sum_{{b=1}\atop{b\neq q}}^n\frac{k_{n+1}\cdot k_b}{(\sigma_{q,b})^2}\delta^{(2)}_q\right).
\end{align*}
The residues of all further terms appearing in (\ref{eq:order}) are computed in exactly the same way. Without showing every single step explicitly, we will give a list of pole orders appearing in the respective terms. Additionally, the results for all residues will be gathered in appendices.

Apart from the computation presented above, the term $\delta^2{\det}_0$ contains one additional expression. It has only first order poles for $\sigma_q\in\{\sigma_1,\sigma_k,\sigma_n\}$, and it has first and second order poles for $\sigma_q\notin\{\sigma_1,\sigma_k,\sigma_n\}$. All residues associated with the term $\delta^2{\det}_0$ are presented in appendix \ref{app:d2d0}.

The residues of the term $\delta^1{\det}_1$ are obtained from three different cases. In the case of $\sigma_q\notin\{\sigma_1,\sigma_2,\sigma_3,\sigma_k,\sigma_n\}$ there are first, second and third order poles. In the case of $\sigma_q\in\{\sigma_1,\sigma_k,\sigma_n\}$ there are first and second order poles. And in the case of $\sigma_q\in\{\sigma_2,\sigma_3\}$ there are first and second order poles. All residues associated with the term $\delta^1{\det}_1$ are presented in appendix \ref{app:d1d1}.

The residues of the term $\delta^0{\det}_2$ are obtained from two different cases. In the case of $\sigma_q\notin\{\sigma_2,\sigma_3\}$ there are first, second and third order poles. And in the case of $\sigma_q\in\{\sigma_2,\sigma_3\}$ there are only first order poles. All residues associated with the term $\delta^0{\det}_2$ are presented in appendix \ref{app:d0d2}.

With this, all relevant terms from higher point expansion are obtained and we can proceed with the computation of lower point construction.

\section{Lower point construction}
\label{sec:lpc}
In the lower point construction we start with the momentum conservation stripped tree level amplitude for $n$ external particles. We set $i=1$, $j=n$, $m=2$ and $w=3$ and invoke the gauge condition $(k_{n+1}\cdot\epsilon_u)=0$ for all $u\in\{1,2,...,n+1\}$, such that\footnote{The gauge condition ensures that there is no remaining $k_{n+1}$ and $\sigma_{n+1}$ dependance in $\det(\Psi')$.}:
\begin{align}
\label{eq:lpamplitude}
M_{n}=\int\left[\prod_{{c=1}\atop{c\neq p,q,r}}^{n}d\sigma_c\right]\frac{4(\sigma_{pq}\sigma_{qr}\sigma_{rp})(\sigma_{1n}\sigma_{nk}\sigma_{k1})}{(\sigma_{2,3})^2}\left[\prod_{{a=2}\atop{a\neq k}}^{n-1}{\delta\left(\sum_{{b=1}\atop{b\neq a}}^{n}\frac{k_a\cdot k_b}{\sigma_{ab}}\right)}\right] \det\left(\Psi'\right),
\end{align}
where we used the abbreviation defined in (\ref{eq:detabbr}). First we notice that only the product of delta functions and the determinant are relevant for our considerations, and all remaining multiplicative factors and integrals are exactly the same spectators which we suppressed in the higher point expansion case. Therefore, here we again suppress these spectator terms, such that the expression we should compare to the higher point expansion is given by:
\begin{align}
\label{eq:lpnospect}
S^{(2)}\left[\prod_{{c=2}\atop{c\neq k}}^{n-1}{\delta\left(\sum_{{b=1}\atop{b\neq c}}^{n}\frac{k_c\cdot k_b}{\sigma_{cb}}\right)}\right] \det\left(\Psi'\right).
\end{align}
As already stated in the introduction, the sub-sub-leading factor $S^{(2)}$ is expected to be given by:
\begin{align}
\label{eq:subsubfactor}
  S^{(2)}=\frac{1}{2}\sum_{a=1}^n\frac{{\epsilon_{n+1}}_{\mu\nu}({k_{n+1}}_\rho J^{\rho\mu}_a)({k_{n+1}}_\lambda J^{\lambda\nu}_a)}{k_{n+1}\cdot k_a},
\end{align}
where the action of the angular momentum operators by their orbital or spin part on momenta or polarization vectors is given by \cite{nima}
\begin{subequations}
\begin{align}
\label{eq:orbitalJ}
J^{\mu\nu}_a k^\beta_b&=\left(k^\mu_a\frac{\partial}{\partial {k_a}_\nu}-k^\nu_a\frac{\partial}{\partial {k_a}_\mu}\right)k^\beta_b \\
\label{eq:spinJ}
J^{\mu\nu}_a \epsilon^\beta_b&=\left(\eta^{\nu\beta}\delta_\sigma^\mu-\eta^{\mu\beta}\delta_\sigma^\nu\right)\epsilon^\sigma_b.
\end{align}
\end{subequations}
Naively, the two angular momentum operators in the sub-sub-leading factor (\ref{eq:subsubfactor}) could act on each other. However, it is trivial to show that the interaction vanishes due to the $(n+1)$\textsuperscript{th} particle being massless $k_{n+1}^2=0$, therefore having only transverse polarization modes $k_{n+1}\cdot\epsilon_{n+1}=0$, and the polarization being light-like such that $\epsilon_{n+1}^2=0$. With this we can conclude that we will have to match the resulting terms to the higher point expansion in the following way
\begin{subequations}
\begin{align}
\label{eq:d2d0}
\frac{1}{2}\sum_{a=1}^n\frac{\epsilon_\mu\epsilon_\nu{q}_\rho {q}_\lambda }{q\cdot k_a}\left(J^{\rho\mu}_a J^{\lambda\nu}_a \prod_{{c=2}\atop{c\neq k}}^{n-1}{\delta\left(\sum_{{b=1}\atop{b\neq c}}^{n}\frac{k_c\cdot k_b}{\sigma_{cb}}\right)}\right) \det\left(\Psi'\right) &~~~\Leftrightarrow~~~ \sum_i \text{Res}_i (\delta^2{\det}_0)\\
\label{eq:d1d1}
\sum_{a=1}^n\frac{\epsilon_\mu\epsilon_\nu{q}_\rho {q}_\lambda }{q\cdot k_a}\left(J^{\rho\mu}_a  \prod_{{c=2}\atop{c\neq k}}^{n-1}{\delta\left(\sum_{{b=1}\atop{b\neq c}}^{n}\frac{k_c\cdot k_b}{\sigma_{cb}}\right)}\right) \left(J^{\lambda\nu}_a \det\left(\Psi'\right)\right) &~~~\Leftrightarrow~~~ \sum_i \text{Res}_i (\delta^1{\det}_1)\\
\label{eq:d0d2}
\frac{1}{2}\sum_{a=1}^n\frac{\epsilon_\mu\epsilon_\nu{q}_\rho {q}_\lambda }{q\cdot k_a} \prod_{{c=2}\atop{c\neq k}}^{n-1}{\delta\left(\sum_{{b=1}\atop{b\neq c}}^{n}\frac{k_c\cdot k_b}{\sigma_{cb}}\right)} \left(J^{\rho\mu}_a J^{\lambda\nu}_a \det\left(\Psi'\right)\right)&~~~\Leftrightarrow~~~ \sum_i \text{Res}_i (\delta^0{\det}_2),
\end{align}
\end{subequations}
where we used the abbreviations 
\begin{align}
\label{eq:eqabbr}
\epsilon_\mu\equiv{\epsilon_{n+1}}_\mu\text{ and }q_\mu\equiv{k_{n+1}}_\mu,
\end{align}
 and the sum in $i$ is over all residues picked up when integrating out $\sigma_{n+1}$.

To compute the lower point construction for (\ref{eq:d2d0}), only the orbital part of the angular momentum operator (\ref{eq:orbitalJ}) is involved, since the scattering equation delta functions depend on momenta only. Therefore, the object of interest is
\begin{align*}
  \frac{1}{2}\sum_{a=1}^n\left((q\cdot k_a)\epsilon^\mu\epsilon^\nu\frac{\partial}{\partial k_a^\mu}\frac{\partial}{\partial k_a^\nu}-2(\epsilon\cdot k_a)\epsilon^\mu q^\lambda\frac{\partial}{\partial k_a^\mu}\frac{\partial}{\partial k_a^\lambda}+\frac{(\epsilon\cdot k_a)^2}{q\cdot k_a}q^\rho q^\lambda\frac{\partial}{\partial k_a^\rho}\frac{\partial}{\partial k_a^\lambda}\right)\prod_{{c=2}\atop{c\neq k}}^{n-1}{\delta\left(\sum_{{b=1}\atop{b\neq c}}^{n}\frac{k_c\cdot k_b}{\sigma_{cb}}\right)}.
\end{align*}
Carrying out the partial derivatives as usual, then suppressing the remaining product of delta functions and abbreviating the derivatives of delta functions in the same way as after (\ref{eq:hpex}), we obtain the same result as from the higher point expansion given in appendix \ref{app:d2d0}. The only type of reshaping needed to recover the exact same set of terms (apart from trivial cancellation and (\ref{eq:theta})), is to combine expressions which have a similar structure up to $\sigma_{ij}$'s appearing in denominators, such that a simplification occurs as in:
\begin{align}
\label{eq:sigmasimp}
\frac{1}{\sigma_{jk}\sigma_{ij}}-\frac{1}{\sigma_{jk}\sigma_{ik}}=\frac{1}{\sigma_{ij}\sigma_{ik}}.
\end{align}
These steps eventually demonstrate the equality of both sides in (\ref{eq:d2d0}).

To compute the lower point construction for (\ref{eq:d1d1}), both parts of the angular momentum operator (\ref{eq:orbitalJ}) and (\ref{eq:spinJ}) are needed. Furthermore, to obtain the derivative of a determinant, we use the chain rule and straightforwardly obtain:
\begin{align}
\label{eq:ddet}
 \frac{d}{dx}\det(A)=\sum_{q=1}^{2(n+1)}\sum_{i=1}^{2(n+1)}(-1)^{q+i}\left(\frac{da_{qi}}{dx}\right)\det(A^i_q).
\end{align}
First, we compute the action of a single angular momentum operator on the product of scattering equation delta functions:
\begin{align}
\label{eq:Jd}
\frac{{\epsilon_{n+1}}_\mu{k_{n+1}}_\rho  }{k_{n+1}\cdot k_a}J^{\rho\mu}_a  \prod_{{c=2}\atop{c\neq k}}^{n-1}{\delta\left(\sum_{{b=1}\atop{b\neq c}}^{n}\frac{k_c\cdot k_b}{\sigma_{cb}}\right)}=& \sum_{{c=2}\atop{c\neq k}}^{n-1}\sum_{{b=1}\atop{b\neq c}}^n\delta_{c,a}\frac{\epsilon_{n+1}\cdot k_b}{\sigma_{cb}}\delta_c^{(1)}+ \sum_{{c=2}\atop{c\neq k}}^{n-1}\sum_{{b=1}\atop{b\neq c}}^n\delta_{b,a}\frac{\epsilon_{n+1}\cdot k_c}{\sigma_{cb}}\delta_c^{(1)} \notag\\
&-\frac{(\epsilon_{n+1}\cdot k_a)}{k_{n+1}\cdot k_a}\sum_{{c=2}\atop{c\neq k}}^{n-1}\sum_{{b=1}\atop{b\neq c}}^n\delta_{c,a}\frac{k_{n+1}\cdot k_b}{\sigma_{cb}}\delta_c^{(1)} \\
&-\frac{(\epsilon_{n+1}\cdot k_a)}{k_{n+1}\cdot k_a}\sum_{{c=2}\atop{c\neq k}}^{n-1}\sum_{{b=1}\atop{b\neq c}}^n\delta_{b,a}\frac{k_{n+1}\cdot k_c}{\sigma_{cb}}\delta_c^{(1)},\notag
\end{align}
where $\delta_{i,j}$ is the Kronecker delta, and where on the right hand side we suppressed the remaining product of delta functions and abbreviated the derivative of the delta function in the same way as after (\ref{eq:hpex}). Next, we compute the action of a single angular momentum operator on the determinant:
\begin{align}
\label{eq:Jdet}
{\epsilon_{n+1}}_\nu {k_{n+1}}_\lambda J^{\lambda\nu}_a\det\left(\Psi'\right)=&
\sum_{{j=1}\atop{j\neq 2,3}}^n \sum_{{i=1}\atop{i\neq 2,3,j}}^n (-1)^{i+j}\left({\epsilon_{n+1}}_\nu {k_{n+1}}_\lambda J^{\lambda\nu}_a\frac{k_j\cdot k_i}{\sigma_{ji}}\right)\det\left(\Psi'^i_j\right) \\
&+\sum_{j=1}^n\sum_{{i=1}\atop{i\neq j}}^n(-1)^{i+j}\left({\epsilon_{n+1}}_\nu {k_{n+1}}_\lambda J^{\lambda\nu}_a\frac{\epsilon_j\cdot \epsilon_i}{\sigma_{ji}}\right)\det\left(\Psi'^{i+n+1}_{j+n+1}\right)\notag\\
&-2\sum_{{j=1}\atop{j\neq 2,3}}^n\sum_{{i=1}\atop{i\neq j}}^n(-1)^{i+j+n+1}\left({\epsilon_{n+1}}_\nu {k_{n+1}}_\lambda J^{\lambda\nu}_a\frac{k_j\cdot \epsilon_i}{\sigma_{ji}}\right)\det\left(\Psi'^{i+n+1}_{j}\right)\notag\\
&-2\sum_{{j=1}\atop{j\neq 2,3}}^n\sum_{{c=1}\atop{c\neq j}}^n(-1)^{n+1}\left({\epsilon_{n+1}}_\nu {k_{n+1}}_\lambda J^{\lambda\nu}_a\frac{k_c\cdot \epsilon_j}{\sigma_{jc}}\right)\det\left(\Psi'^{j+n+1}_{j}\right).\notag
\end{align}
To make the terms more explicit, we invoke the usual gauge from before $k_{n+1}\cdot \epsilon_i=0$ for all $i$ and obtain:
\begin{subequations}
\label{eq:Jterms}
\begin{align}
{\epsilon_{n+1}}_\nu {k_{n+1}}_\lambda J^{\lambda\nu}_a\frac{k_j\cdot k_i}{\sigma_{ji}}=&(\delta_{a,j}-\delta_{a,i})\frac{(k_{n+1}\cdot k_j)(\epsilon_{n+1}\cdot k_i)-(k_{n+1}\cdot k_i)(\epsilon_{n+1}\cdot k_j)}{\sigma_{ji}} \\
{\epsilon_{n+1}}_\nu {k_{n+1}}_\lambda J^{\lambda\nu}_a\frac{ k_j\cdot\epsilon_i}{\sigma_{ji}}=&(\delta_{a,j}-\delta_{a,i})\frac{(k_{n+1}\cdot k_j)(\epsilon_{n+1}\cdot \epsilon_i)}{\sigma_{ji}} \\
{\epsilon_{n+1}}_\nu {k_{n+1}}_\lambda J^{\lambda\nu}_a\frac{ \epsilon_j\cdot\epsilon_i}{\sigma_{ji}}=&\,0.
\end{align}
\end{subequations}
Plugging (\ref{eq:Jterms}) into (\ref{eq:Jdet}), multiplying with (\ref{eq:Jd}) and summing over $a=1,...,n$ gives the same result as from higher point expansion in appendix \ref{app:d1d1}. To recover the exact same set of terms in order to prove the equality, we use simplifications like (\ref{eq:theta}) and (\ref{eq:sigmasimp}). Additionally, we realize that for an antisymmetric $2(n+1)\times 2(n+1)$ matrix $A$ we have:
\begin{align}
\label{eq:dett}
\det\left(A^{a_1,a_2,...,a_m}_{b_1,b_2,...,b_m}\right)=(-1)^{m}\det\left(A^{b_1,b_2,...,b_m}_{a_1,a_2,...,a_m}\right).
\end{align}
Making use of these steps, the demonstration of the equality of both sides in (\ref{eq:d1d1}) becomes straightforward.

Finally, to compute the lower point construction for (\ref{eq:d0d2}), again both parts of the angular momentum operator (\ref{eq:orbitalJ}) and (\ref{eq:spinJ}) are needed. We start with (\ref{eq:Jdet}) and act with the angular momentum operator a second time. The case where both angular momentum operators hit the expansion coefficient in each line vanishes due to the same arguments as the vanishing of the self-interaction of the two angular momentum operators. Therefore, only the case remains where the second angular momentum operator acts on the determinant in each line. Combining (\ref{eq:Jdet}) with (\ref{eq:Jterms}) and using the abbreviations (\ref{eq:eqabbr}), this results in:
\begin{align}
\label{eq:JJdet}
\frac{1}{2}\sum_{a=1}^n\frac{\epsilon_\mu\epsilon_\nu q_\rho q_\lambda}{q\cdot k_a}J^{\rho\mu}_a J^{\lambda\nu}_a\det\left(\Psi'\right)
=&\sum_{{m=1}\atop{m\neq 2,3}}^n\sum_{{i=1}\atop{i\neq 2,3,m}}^n (-1)^{i+m}\frac{(\epsilon\cdot k_i)}{\sigma_{mi}}\left(q_\rho\epsilon_\mu J^{\rho\mu}_m\det\left(\Psi'^i_m\right)\right) \\
&-\sum_{{m=1}\atop{m\neq 2,3}}^n\sum_{{i=1}\atop{i\neq 2,3,m}}^n (-1)^{i+m}\frac{(\epsilon\cdot k_m)(q\cdot k_i)}{(q\cdot k_m)\sigma_{mi}}\left(q_\rho\epsilon_\mu J^{\rho\mu}_m\det\left(\Psi'^i_m\right)\right)\notag \\
&+\sum_{{m=1}\atop{m\neq 2,3}}^n\sum_{{i=1}\atop{i\neq m}}^n (-1)^{i+m+n+1}\frac{(\epsilon\cdot \epsilon_i)(q\cdot k_m)}{(q\cdot k_i)\sigma_{mi}}\left(q_\rho\epsilon_\mu J^{\rho\mu}_i\det\left(\Psi'^{i+n+1}_m\right)\right)\notag \\
&-\sum_{{m=1}\atop{m\neq 2,3}}^n\sum_{{i=1}\atop{i\neq m}}^n (-1)^{i+m+n+1}\frac{(\epsilon\cdot \epsilon_i)}{\sigma_{mi}}\left(q_\rho\epsilon_\mu J^{\rho\mu}_m\det\left(\Psi'^{i+n+1}_m\right)\right)\notag \\
&+\sum_{{m=1}\atop{m\neq 2,3}}^n\sum_{{i=1}\atop{i\neq m}}^n (-1)^{n+1}\frac{(\epsilon\cdot \epsilon_m)(q\cdot k_i)}{(q\cdot k_m)\sigma_{mi}}\left(q_\rho\epsilon_\mu J^{\rho\mu}_m\det\left(\Psi'^{m+n+1}_m\right)\right)\notag \\
&-\sum_{{m=1}\atop{m\neq 2,3}}^n\sum_{{i=1}\atop{i\neq m}}^n (-1)^{n+1}\frac{(\epsilon\cdot \epsilon_m)}{\sigma_{mi}}\left(q_\rho\epsilon_\mu J^{\rho\mu}_i\det\left(\Psi'^{m+n+1}_m\right)\right).\notag
\end{align}
The action of the angular momentum operator on the determinants in each of these six lines is then expanded further analogously to (\ref{eq:Jdet}). The only difference is, that now the expansion summations have to omit one removed row and column more in each case, and we have to explicitly display the corresponding step functions in the exponent of (-1). Since the product of scattering equation delta functions is untouched by the operators in this case, it can be suppressed as a spectator completely, so that the terms resulting from a further expansion of (\ref{eq:JJdet}) correspond to the higher point expansion result given in appendix \ref{app:d0d2}. Again, making use of simplifications (\ref{eq:theta}), (\ref{eq:sigmasimp}) and (\ref{eq:dett}), it is then straightforward to reshape the finding to obtain the exact same set of terms listed in appendix \ref{app:d0d2}, which proves the equality of both sides in (\ref{eq:d0d2}).

This concludes the computation of the lower point construction and its comparison with the higher point expansion. Both yield the same result, which confirms that the sub-sub-leading factor (\ref{eq:subsublead}) in the soft-graviton expansion of tree level scattering amplitudes is indeed valid in arbitrary dimension.

\newpage
\appendix
\section{Residues of $\delta^2{\det}_0$}
\label{app:d2d0}
The following are all residues obtained from $\delta^2{\det}_0$ in (\ref{eq:order}) by integrating out the $\sigma_{n+1}$ dependance. Multiplicative spectator terms and integrals which are trivially the same in the lower point construction are suppressed.\footnote{In this particular case the determinant $\det(\Psi')$ is also suppressed, since it is also a multiplicative spectator term in $\delta^2{\det}_0$.} Additionally, the product of scattering equation delta functions is suppressed and the derivative of delta function is abbreviated as 
\begin{align}
\label{eq:deltaabbr}
\delta^{(i)}_j\equiv\delta^{(i)}\left(\sum_{{a=1}\atop{a\neq j}}^n\frac{k_a\cdot k_j}{\sigma_{aj}}\right).
\end{align}
With this the residues are:
\begin{align}
&2\sum_{{q=2}\atop{q\neq k}}^{n-1}(\epsilon_{n+1}\cdot k_q)\delta^{(1)}_q\sum_{{r=2}\atop{r\neq k,q}}^{n-1}\frac{k_r\cdot k_{n+1}}{\sigma_{qr}}\delta^{(1)}_r\sum_{{b=1}\atop{b\neq q}}^n\frac{\epsilon_{n+1}\cdot k_b}{\sigma_{qb}} \notag\\
&+\frac{1}{2}\sum_{{r=2}\atop{r\neq k}}^{n-1} (k_r\cdot k_{n+1})\delta^{(1)}_r\sum_{{q=1}\atop{q\neq r}}^n\frac{1}{k_{n+1}\cdot k_q}\frac{(\epsilon_{n+1}\cdot k_q)^2}{\sigma_{qr}}\sum_{{t=2}\atop{t\neq k,r,q}}^{n-1} \frac{k_t\cdot k_{n+1}}{\sigma_{qt}}\delta^{(1)}_t \notag\\
&-\sum_{{q=2}\atop{q\neq k}}^{n-1}(\epsilon_{n+1}\cdot k_q)^2\delta^{(1)}_q\left(\sum_{{r=2}\atop{r\neq k,q}}^{n-1}\frac{k_r\cdot k_{n+1}}{(\sigma_{qr})^2}\delta^{(1)}_r+\frac{1}{k_{n+1}\cdot k_q}\sum_{{r=2}\atop{r\neq k,q}}^{n-1}\frac{k_r\cdot k_{n+1}}{\sigma_{qr}}\delta^{(1)}_r\sum_{{b=1}\atop{b\neq q}}^n\frac{k_{n+1}\cdot k_b}{\sigma_{qb}}\right) \notag \\
&+\frac{1}{2}\sum_{{q=2}\atop{q\neq k}}^{n-1}(k_q\cdot k_{n+1})\left(\sum_{{b=1}\atop{b\neq q}}^{n}\frac{\epsilon_{n+1}\cdot k_c}{\sigma_{q,c}}\right)^2\delta^{(2)}_q+\frac{1}{2}\sum_{{q=1}}^{n}\frac{(\epsilon_{n+1}\cdot k_q)^2}{k_{n+1}\cdot k_q}\sum_{{r=2}\atop{r\neq k,q}}^{n-1}\left(\frac{k_r\cdot k_{n+1}}{\sigma_{q,r}}\right)^2\delta^{(2)}_r \notag\\
&-\sum_{{q=2}\atop{q\neq k}}^{n-1}(\epsilon_{n+1}\cdot k_q)\left((k_q\cdot k_{n+1})\sum_{{b=1}\atop{b\neq q}}^n\frac{\epsilon_{n+1}\cdot k_b}{(\sigma_{q,b})^2}\delta^{(2)}_q+\sum_{{b=1}\atop{b\neq q}}^n\frac{\epsilon_{n+1}\cdot k_b}{\sigma_{q,b}}\sum_{{c=1}\atop{c\neq q}}^n\frac{k_{n+1}\cdot k_c}{\sigma_{q,c}}\delta^{(2)}_q\right) \notag\\
&+\frac{1}{2}\sum_{{q=2}\atop{q\neq k}}^{n-1}(\epsilon_{n+1}\cdot k_q)^2\left(\frac{1}{k_{n+1}\cdot k_q}\left(\sum_{{b=1}\atop{b\neq q}}^n\frac{k_{n+1}\cdot k_b}{\sigma_{q,b}}\right)^2\delta^{(2)}_q+\sum_{{b=1}\atop{b\neq q}}^n\frac{k_{n+1}\cdot k_b}{(\sigma_{q,b})^2}\delta^{(2)}_q\right).\notag
\end{align}

\section{Residues of $\delta^1{\det}_1$}
\label{app:d1d1}
The following are all residues obtained from $\delta^1{\det}_1$ in (\ref{eq:order}) by integrating out the $\sigma_{n+1}$ dependance. Multiplicative spectator terms and integrals which are trivially the same in the lower point construction are suppressed. Additionally, the product of scattering equation delta functions is suppressed and the derivative of delta function is abbreviated as (\ref{eq:deltaabbr}). With this the residues are:
\begin{align*}
&-2\sum_{{q=2}\atop{q\neq k}}^{n-1}(\epsilon_{n+1}\cdot k_q)\delta^{(1)}_q\sum_{{i=1}\atop{i\neq 2,3,q}}^n\frac{k_{n+1}\cdot k_i}{\sigma_{qi}}\sum_{{j=1}\atop{j\neq 2,3,q}}^n(-1)^{i+j}\frac{\epsilon_{n+1}\cdot k_j}{\sigma_{qj}}\det\left(\Psi'^{i}_j\right)\\
&-2\sum_{{q=4}\atop{q\neq k}}^{n-1}(k_{n+1}\cdot k_q)\delta^{(1)}_{q}\sum_{{c=1}\atop{c\neq q}}^{n}\frac{\epsilon_{n+1}\cdot k_{c}}{\sigma_{qc}}\sum_{{j=1}\atop{j\neq 2,3,q}}^{n}(-1)^{q+j}\frac{\epsilon_{n+1}\cdot k_{j}}{\sigma_{qj}}\det\left(\Psi'^{q}_{j}\right)\\
&-2\sum_{{q=4}\atop{q\neq k}}^{n-1}(\epsilon_{n+1}\cdot k_{q})\delta^{(1)}_{q}\sum_{{c=1}\atop{c\neq q}}^{n}\frac{\epsilon_{n+1}\cdot k_{c}}{\sigma_{qc}}\sum_{{i=1}\atop{i\neq 2,3,q}}^{n}(-1)^{i+q}\frac{k_{n+1}\cdot k_{i}}{\sigma_{qi}}\det\left(\Psi'^{i}_{q}\right)\\
&+2\sum_{{q=4}\atop{q\neq k}}^{n-1}(\epsilon_{n+1}\cdot k_{q})\delta^{(1)}_{q}\sum_{{b=1}\atop{b\neq q}}^{n}\frac{k_{n+1}\cdot k_{b}}{\sigma_{qb}}\sum_{{j=1}\atop{j\neq 2,3,q}}^{n}(-1)^{q+j}\frac{\epsilon_{n+1}\cdot k_{j}}{\sigma_{qj}}\det\left(\Psi'^{q}_{j}\right)\\
&+2\sum_{{q=4}\atop{q\neq k}}^{n-1}\frac{(\epsilon_{n+1}\cdot k_{q})^2}{k_{n+1}\cdot k_q}\delta^{(1)}_{q}\sum_{{b=1}\atop{b\neq q}}^{n}\frac{k_{n+1}\cdot k_{b}}{\sigma_{qb}}\sum_{{i=1}\atop{i\neq 2,3,q}}^{n}(-1)^{i+q}\frac{k_{n+1}\cdot k_{i}}{\sigma_{qi}}\det\left(\Psi'^{i}_{q}\right)\\
&-2\sum_{{q=1}\atop{q\neq 2,3}}^{n}(\epsilon_{n+1}\cdot k_{q})\sum_{{r=2}\atop{r\neq k,q}}^{n-1}\frac{k_{n+1}\cdot k_{r}}{\sigma_{qr}}\delta^{(1)}_{r}\sum_{{j=1}\atop{j\neq 2,3,q}}^{n}(-1)^{q+j}\frac{\epsilon_{n+1}\cdot k_{j}}{\sigma_{qj}}\det\left(\Psi'^{q}_{j}\right)\\
&-2\sum_{{q=1}\atop{q\neq 2,3}}^{n}\frac{(\epsilon_{n+1}\cdot k_{q})^2}{(k_{n+1}\cdot k_{q})}\sum_{{r=2}\atop{r\neq k,q}}^{n-1}\frac{k_{n+1}\cdot k_{r}}{\sigma_{qr}}\delta^{(1)}_{r}\sum_{{i=1}\atop{i\neq 2,3,q}}^{n}(-1)^{i+q}\frac{k_{n+1}\cdot k_{i}}{\sigma_{qi}}\det\left(\Psi'^{i}_{q}\right)\\
&+2\sum_{{q=4}\atop{q\neq k}}^{n-1}(\epsilon_{n+1}\cdot k_{q})(k_{n+1}\cdot k_{q})\delta^{(1)}_{q}\sum_{{j=1}\atop{j\neq 2,3,q}}^{n}(-1)^{q+j}\frac{\epsilon_{n+1}\cdot k_{j}}{(\sigma_{qj})^2}\det\left(\Psi'^{q}_{j}\right)\\
&+2\sum_{{q=4}\atop{q\neq k}}^{n-1}(\epsilon_{n+1}\cdot k_{})^2\delta^{(1)}_{q}\sum_{{i=1}\atop{i\neq 2,3,q}}^{n}(-1)^{i+q}\frac{k_{n+1}\cdot k_{i}}{(\sigma_{qi})^2}\det\left(\Psi'^{i}_{q}\right)\\
&+2\sum_{{q=2}\atop{q\neq k}}^{n-1}(\epsilon_{n+1}\cdot k_{q})\delta^{(1)}_{q}\sum_{{i=1}\atop{i\neq 2,3,q}}^{n}\frac{k_{n+1}\cdot k_{i}}{\sigma_{qi}}\sum_{{j=1}\atop{j\neq q}}^{n}(-1)^{i+j+n+1}\frac{\epsilon_{n+1}\cdot \epsilon_{j}}{\sigma_{qj}}\det\left(\Psi'^{i}_{j+n+1}\right)\\
&+2\sum_{{q=4}\atop{q\neq k}}^{n-1}k_{n+1}\cdot k_{q}\delta^{(1)}_{q}\sum_{{c=1}\atop{c\neq q}}^{n}\frac{\epsilon_{n+1}\cdot k_{c}}{\sigma_{qc}}\sum_{{j=1}\atop{j\neq q}}^{n}(-1)^{q+j+n+1}\frac{\epsilon_{n+1}\cdot \epsilon_{j}}{\sigma_{qj}}\det\left(\Psi'^{q}_{j+n+1}\right)\\
&+2\sum_{{q=2}\atop{q\neq k}}^{n-1}(\epsilon_{n+1}\cdot \epsilon_{q})\delta^{(1)}_{q}\sum_{{c=1}\atop{c\neq q}}^{n}\frac{\epsilon_{n+1}\cdot k_{c}}{\sigma_{qc}}\sum_{{i=1}\atop{i\neq 2,3,q}}^{n}(-1)^{i+q+n+1}\frac{k_{n+1}\cdot k_{i}}{\sigma_{qi}}\det\left(\Psi'^{i}_{q+n+1}\right)\\
&-2\sum_{{q=4}\atop{q\neq k}}^{n-1}(\epsilon_{n+1}\cdot k_{q})\delta^{(1)}_{q}\sum_{{b=1}\atop{b\neq q}}^{n}\frac{k_{n+1}\cdot k_{b}}{\sigma_{qb}}\sum_{{j=1}\atop{j\neq q}}^{n}(-1)^{q+j+n+1}\frac{\epsilon_{n+1}\cdot \epsilon_{j}}{\sigma_{qj}}\det\left(\Psi'^{q}_{j+n+1}\right)\\
&-2\sum_{{q=2}\atop{q\neq k}}^{n-1}\frac{(\epsilon_{n+1}\cdot k_{q})(\epsilon_{n+1}\cdot \epsilon_{q})}{(k_{n+1}\cdot k_{q})}\delta^{(1)}_{q}\sum_{{b=1}\atop{b\neq q}}^{n}\frac{k_{n+1}\cdot k_{b}}{\sigma_{qb}}\sum_{{i=1}\atop{i\neq 2,3,q}}^{n}(-1)^{i+q+n+1}\frac{k_{n+1}\cdot k_{i}}{\sigma_{qi}}\det\left(\Psi'^{i}_{q+n+1}\right)
\end{align*}
\begin{align*}
&+2\sum_{{q=1}\atop{q\neq 2,3}}^{n}(\epsilon_{n+1}\cdot k_{q})\sum_{{r=2}\atop{r\neq k,q}}^{n-1}\frac{k_{n+1}\cdot k_{r}}{\sigma_{qr}}\delta^{(1)}_{r}\sum_{{j=1}\atop{j\neq q}}^{n}(-1)^{q+j+n+1}\frac{\epsilon_{n+1}\cdot \epsilon_{j}}{\sigma_{qj}}\det\left(\Psi'^{q}_{j+n+1}\right)\\
&+2\sum_{{q=1}}^{n}\frac{(\epsilon_{n+1}\cdot k_{q})(\epsilon_{n+1}\cdot \epsilon_{q})}{(k_{n+1}\cdot k_{q})}\sum_{{r=2}\atop{r\neq k,q}}^{n-1}\frac{k_{n+1}\cdot k_{r}}{\sigma_{qr}}\delta^{(1)}_{r}\sum_{{i=1}\atop{i\neq 2,3,q}}^{n}(-1)^{i+q+n+1}\frac{k_{n+1}\cdot k_{i}}{\sigma_{qi}}\det\left(\Psi'^{i}_{q+n+1}\right)\\
&-2\sum_{{q=4}\atop{q\neq k}}^{n-1}(\epsilon_{n+1}\cdot k_{q})(k_{n+1}\cdot k_{q})\delta^{(1)}_{q}\sum_{{j=1}\atop{j\neq q}}^{n}(-1)^{q+j+n+1}\frac{\epsilon_{n+1}\cdot \epsilon_{j}}{(\sigma_{qj})^2}\det\left(\Psi'^{q}_{j+n+1}\right)\\
&-2\sum_{{q=2}\atop{q\neq k}}^{n-1}(\epsilon_{n+1}\cdot k_{q})(\epsilon_{n+1}\cdot \epsilon_{q})\delta^{(1)}_{q}\sum_{{i=1}\atop{i\neq 2,3,q}}^{n}(-1)^{i+q+n+1}\frac{k_{n+1}\cdot k_{i}}{(\sigma_{qi})^2}\det\left(\Psi'^{i}_{q+n+1}\right)\\
&-2\sum_{{q=4}\atop{q\neq k}}^{n-1}(\epsilon_{n+1}\cdot \epsilon_{q})(k_{n+1}\cdot k_{q})\delta^{(1)}_{q}\sum_{{c=1}\atop{c\neq q}}^{n}(-1)^{n+1}\frac{\epsilon_{n+1}\cdot k_{c}}{(\sigma_{qc})^2}\det\left(\Psi'^{q}_{q+n+1}\right)\\
&-2\sum_{{q=4}\atop{q\neq k}}^{n-1}(\epsilon_{n+1}\cdot \epsilon_{q})\delta^{(1)}_{q}\sum_{{c=1}\atop{c\neq q}}^{n}\frac{\epsilon_{n+1}\cdot k_{c}}{\sigma_{qc}}\sum_{{b=1}\atop{b\neq q}}^{n}(-1)^{n+1}\frac{k_{n+1}\cdot k_{b}}{\sigma_{qb}}\det\left(\Psi'^{q}_{q+n+1}\right)\\
&+2\sum_{{q=1}\atop{q\neq 2,3}}^{n}(\epsilon_{n+1}\cdot \epsilon_{q})\sum_{{c=1}\atop{c\neq q}}^{n}\frac{\epsilon_{n+1}\cdot k_{c}}{\sigma_{qc}}\sum_{{r=2}\atop{r\neq k,q}}^{n-1}(-1)^{n+1}\frac{k_{n+1}\cdot k_{r}}{\sigma_{qr}}\delta^{(1)}_{r}\det\left(\Psi'^{q}_{q+n+1}\right)\\
&+2\sum_{{q=4}\atop{q\neq k}}^{n-1}\frac{(\epsilon_{n+1}\cdot k_{q})(\epsilon_{n+1}\cdot \epsilon_{q})}{(k_{n+1}\cdot k_{q})}\delta^{(1)}_{q}\left(\sum_{{b=1}\atop{b\neq q}}^{n}\frac{k_{n+1}\cdot k_{b}}{\sigma_{qb}}\right)^2(-1)^{n+1}\det\left(\Psi'^{q}_{q+n+1}\right)\\
&+2\sum_{{q=4}\atop{q\neq k}}^{n-1}(\epsilon_{n+1}\cdot k_{q})(\epsilon_{n+1}\cdot \epsilon_{q})\delta^{(1)}_{q}\sum_{{b=1}\atop{b\neq q}}^{n}(-1)^{n+1}\frac{k_{n+1}\cdot k_{b}}{(\sigma_{qb})^2}\det\left(\Psi'^{q}_{q+n+1}\right)\\
&-2\sum_{{q=1}\atop{q\neq 2,3}}^{n}\frac{(\epsilon_{n+1}\cdot k_{q})(\epsilon_{n+1}\cdot \epsilon_{q})}{(k_{n+1}\cdot k_{q})}\sum_{{b=1}\atop{b\neq q}}^{n}\frac{k_{n+1}\cdot k_{b}}{\sigma_{qb}}\sum_{{r=2}\atop{r\neq k,q}}^{n-1}(-1)^{n+1}\frac{k_{n+1}\cdot k_{r}}{\sigma_{qr}}\delta^{(1)}_{r}\det\left(\Psi'^{q}_{q+n+1}\right)\\
&-2\sum_{{q=1}\atop{q\neq 2,3}}^{n}(\epsilon_{n+1}\cdot k_{q})(\epsilon_{n+1}\cdot \epsilon_{q})\sum_{{r=2}\atop{r\neq k,q}}^{n-1}(-1)^{n+1}\frac{k_{n+1}\cdot k_{r}}{(\sigma_{qr})^2}\delta^{(1)}_{r}\det\left(\Psi'^{q}_{q+n+1}\right)
\end{align*}

\section{Residues of $\delta^0{\det}_2$}
\label{app:d0d2}
The following are all residues obtained from $\delta^0{\det}_2$ in (\ref{eq:order}) by integrating out the $\sigma_{n+1}$ dependance. Multiplicative spectator terms and integrals which are trivially the same in the lower point construction are suppressed. Additionally, the product of scattering equation delta functions is suppressed. With this the residues are:
\begin{align*}
&\sum_{{q=1}\atop{q\neq 2,3}}^{n}(k_{n+1}\cdot k_{q})\sum_{{u=1}\atop{u\neq 2,3,q}}^{n}\sum_{{j=1}\atop{j\neq 2,3,q}}^{n}(-1)^{u+j+\theta(u,q)+\theta(j,q)}\frac{\epsilon_{n+1}\cdot k_{u}}{\sigma_{qu}}\frac{\epsilon_{n+1}\cdot k_{j}}{\sigma_{qj}}\det\left(\Psi'^{q,u}_{q,j}\right)\\
&+2\sum_{{q=1}\atop{q\neq 2,3}}^{n}(\epsilon_{n+1}\cdot k_{q})\sum_{{j=1}\atop{j\neq 2,3,q}}^{n}\sum_{{u=1}\atop{u\neq 2,3,q}}^{n}(-1)^{j+u+\theta(u,q)+\theta(q,j)}\frac{k_{n+1}\cdot k_{j}}{\sigma_{qj}}\frac{\epsilon_{n+1}\cdot k_{u}}{\sigma_{qu}}\det\left(\Psi'^{q,u}_{j,q}\right)\\
&+\sum_{{q=1}\atop{q\neq 2,3}}^{n}\frac{(\epsilon_{n+1}\cdot \epsilon_{q})^2}{(k_{n+1}\cdot k_{q})}\sum_{{i=1}\atop{i\neq 2,3,q}}^{n}\sum_{{j=1}\atop{j\neq 2,3,q}}^{n}(-1)^{i+j+\theta(q,i)+\theta(q,j)}\frac{k_{n+1}\cdot k_{i}}{\sigma_{qi}}\frac{k_{n+1}\cdot k_{j}}{\sigma_{qj}}\det\left(\Psi'^{i,q}_{j,q}\right)\\
&-2\sum_{{q=1}\atop{q\neq 2,3}}^{n}(\epsilon_{n+1}\cdot \epsilon_{q})\sum_{{j=1}\atop{j\neq 2,3,q}}^{n}\sum_{{u=1}\atop{u\neq 2,3,q}}^{n}(-1)^{j+u+n+1+\theta(u,q)+\theta(q+n+1,j)}\frac{k_{n+1}\cdot k_{j}}{\sigma_{qj}}\frac{\epsilon_{n+1}\cdot k_{u}}{\sigma_{qu}}\det\left(\Psi'^{q,u}_{j,q+n+1}\right)\\
&-2\sum_{{q=1}\atop{q\neq 2,3}}^{n}(\epsilon_{n+1}\cdot \epsilon_{q})\sum_{{i=1}\atop{i\neq 2,3,q}}^{n}\sum_{{u=1}\atop{u\neq 2,3,q,i}}^{n}(-1)^{i+u+n+1+\theta(u,i)+\theta(q+n+1,q)}\frac{k_{n+1}\cdot k_{i}}{\sigma_{qi}}\frac{\epsilon_{n+1}\cdot k_{u}}{\sigma_{qu}}\det\left(\Psi'^{i,u}_{q,q+n+1}\right)\\
&+2\sum_{{q=1}\atop{q\neq 2,3}}^{n}(\epsilon_{n+1}\cdot \epsilon_{q})\sum_{{b=1}\atop{b\neq q}}^{n}\frac{k_{n+1}\cdot k_{b}}{\sigma_{qb}}\sum_{{u=1}\atop{u\neq 2,3,q}}^{n}(-1)^{q+u+n+1+\theta(u,q)+\theta(q+n+1,q)}\frac{\epsilon_{n+1}\cdot k_{u}}{\sigma_{qu}}\det\left(\Psi'^{q,u}_{q,q+n+1}\right)\\
&-2\sum_{{q=1}\atop{q\neq 2,3}}^{n}\frac{(\epsilon_{n+1}\cdot k_{q})(\epsilon_{n+1}\cdot \epsilon_{q})}{(k_{n+1}\cdot k_{q})}\sum_{{i=1}\atop{i\neq 2,3,q}}^{n}\sum_{{j=1}\atop{j\neq 2,3,q}}^{n}(-1)^{i+j+n+1+\theta(q,i)+\theta(q+n+1,j)}\frac{k_{n+1}\cdot k_{i}}{\sigma_{qi}}\frac{k_{n+1}\cdot k_{j}}{\sigma_{qj}}\det\left(\Psi'^{i,q}_{j,q+n+1}\right)\\
&-2\sum_{{q=1}\atop{q\neq 2,3}}^{n}(k_{n+1}\cdot k_{q})\sum_{{u=1}\atop{u\neq 2,3,q}}^{n}\sum_{{j=1}\atop{j\neq q}}^{n}(-1)^{u+j+n+1+\theta(u,q)+\theta(j+n+1,q)}\frac{\epsilon_{n+1}\cdot k_{u}}{\sigma_{qu}}\frac{\epsilon_{n+1}\cdot \epsilon_{j}}{\sigma_{qj}}\det\left(\Psi'^{q,u}_{q,j+n+1}\right)\\
&-2\sum_{{q=1}\atop{q\neq 2,3}}^{n}(\epsilon_{n+1}\cdot k_{q})\sum_{{i=1}\atop{i\neq 2,3,q}}^{n}\sum_{{j=1}\atop{j\neq q}}^{n}(-1)^{i+j+n+1+\theta(q,i)+\theta(j+n+1,q)}\frac{k_{n+1}\cdot k_{i}}{\sigma_{qi}}\frac{\epsilon_{n+1}\cdot \epsilon_{j}}{\sigma_{qj}}\det\left(\Psi'^{i,q}_{q,j+n+1}\right)\\
&+2\sum_{{q=1}\atop{q\neq 2,3}}^{n}\frac{(\epsilon_{n+1}\cdot k_{q})(\epsilon_{n+1}\cdot \epsilon_{q})}{(k_{n+1}\cdot k_{q})}\sum_{{b=1}\atop{b\neq q}}^{n}\frac{k_{n+1}\cdot k_{b}}{\sigma_{qb}}\sum_{{i=1}\atop{i\neq 2,3,q}}^{n}(-1)^{i+q+n+1+\theta(q,i)+\theta(q+n+1,q)}\frac{k_{n+1}\cdot k_{i}}{\sigma_{qi}}\det\left(\Psi'^{i,q}_{q,q+n+1}\right)\\
&+2\sum_{{q=1}\atop{q\neq 2,3}}^{n}(k_{n+1}\cdot k_{q})(\epsilon_{n+1}\cdot \epsilon_{q})\sum_{{u=1}\atop{u\neq 2,3,q}}^{n}(-1)^{q+u+n+1+\theta(u,q)+\theta(q+n+1,q)}\frac{\epsilon_{n+1}\cdot k_{u}}{(\sigma_{qu})^2}\det\left(\Psi'^{q,u}_{q,q+n+1}\right)\\
&+2\sum_{{q=1}\atop{q\neq 2,3}}^{n}(\epsilon_{n+1}\cdot \epsilon_{q})(\epsilon_{n+1}\cdot k_{q})\sum_{{i=1}\atop{i\neq 2,3,q}}^{n}(-1)^{i+q+n+1+\theta(q,i)+\theta(q+n+1,q)}\frac{k_{n+1}\cdot k_{i}}{\sigma_{qi}}\det\left(\Psi'^{i,q}_{q,q+n+1}\right)\\
&+\sum_{{q=1}\atop{q\neq 2,3}}^{n}(\epsilon_{n+1}\cdot \epsilon_{q})^2\sum_{{b=1}\atop{b\neq q}}^{n}\frac{k_{n+1}\cdot k_{b}}{(\sigma_{qb})^2}(-1)^{\theta(q+n+1,q)+\theta(q+n+1,q)}\det\left(\Psi'^{q,q+n+1}_{q,q+n+1}\right)\\
&+\sum_{{q=1}\atop{q\neq 2,3}}^{n}\frac{(\epsilon_{n+1}\cdot \epsilon_{q})^2}{(k_{n+1}\cdot k_{q})}\left(\sum_{{b=1}\atop{b\neq q}}^{n}\frac{k_{n+1}\cdot k_{b}}{\sigma_{qb}}\right)^2 (-1)^{\theta(q+n+1,q)+\theta(q+n+1,q)}\det\left(\Psi'^{q,q+n+1}_{q,q+n+1}\right)
\end{align*}
\begin{align*}
&-2\sum_{{q=1}\atop{q\neq 2,3}}^{n}\frac{(\epsilon_{n+1}\cdot \epsilon_{q})^2}{(k_{n+1}\cdot k_{q})}\sum_{{b=1}\atop{b\neq q}}^{n}\sum_{{i=1}\atop{i\neq 2,3,q}}^{n}(-1)^{i+q+\theta(q+n+1,i)+\theta(q+n+1,q)}\frac{k_{n+1}\cdot k_{i}}{\sigma_{qi}}\det\left(\Psi'^{i,q+n+1}_{q,q+n+1}\right)\\
&-2\sum_{{q=1}\atop{q\neq 2,3}}^{n}(\epsilon_{n+1}\cdot \epsilon_{q})\sum_{{b=1}\atop{b\neq q}}^{n}\frac{k_{n+1}\cdot k_{b}}{\sigma_{qb}}\sum_{{u=1}\atop{u\neq q}}^{n}(-1)^{u+q+\theta(u+n+1,q)+\theta(q+n+1,q)}\frac{\epsilon_{n+1}\cdot \epsilon_{u}}{\sigma_{qu}}\det\left(\Psi'^{q,u+n+1}_{q,q+n+1}\right)\\
&-2\sum_{{q=1}\atop{q\neq 2,3}}^{n}(\epsilon_{n+1}\cdot \epsilon_{q})^2\sum_{{i=1}\atop{i\neq 2,3,q}}^{n}(-1)^{i+q+\theta(q+n+1,i)+\theta(q+n+1,q)}\frac{k_{n+1}\cdot k_{i}}{(\sigma_{qi})^2}\det\left(\Psi'^{i,q+n+1}_{q,q+n+1}\right)\\
&-2\sum_{{q=1}\atop{q\neq 2,3}}^{n}(k_{n+1}\cdot k_{q})(\epsilon_{n+1}\cdot \epsilon_{q})\sum_{{u=1}\atop{u\neq q}}^{n}(-1)^{u+q+\theta(u+n+1,q)+\theta(q+n+1,q)}\frac{\epsilon_{n+1}\cdot \epsilon_{u}}{(\sigma_{qu})^2}\det\left(\Psi'^{q,u+n+1}_{q,q+n+1}\right)\\
&+2\sum_{{q=1}\atop{q\neq 2,3}}^{n}(\epsilon_{n+1}\cdot \epsilon_{q})\sum_{{i=1}\atop{i\neq 2,3,q}}^{n}\sum_{{u=1}\atop{u\neq q}}^{n}(-1)^{i+u+\theta(u+n+1,i)+\theta(q+n+1,q)}\frac{k_{n+1}\cdot k_{i}}{\sigma_{qi}}\frac{\epsilon_{n+1}\cdot \epsilon_{u}}{\sigma_{qu}}\det\left(\Psi'^{q,q+n+1}_{i,u+n+1}\right)\\
&+2\sum_{{q=1}\atop{q\neq 2,3}}^{n}(\epsilon_{n+1}\cdot \epsilon_{q})\sum_{{i=1}\atop{i\neq 2,3,q}}^{n}\sum_{{u=1}\atop{u\neq q}}^{n}(-1)^{i+u+\theta(u+n+1,i)+\theta(q+n+1,q)}\frac{k_{n+1}\cdot k_{i}}{\sigma_{qi}}\frac{\epsilon_{n+1}\cdot \epsilon_{u}}{\sigma_{qu}}\det\left(\Psi'^{i,q+n+1}_{q,u+n+1}\right)\\
&+\sum_{{q=1}\atop{q\neq 2,3}}^{n}(k_{n+1}\cdot k_{q})\sum_{{u=1}\atop{u\neq q}}^{n}\sum_{{j=1}\atop{j\neq q}}^{n}(-1)^{u+j+\theta(u+n+1,q)+\theta(j+n+1,q)}\frac{\epsilon_{n+1}\cdot \epsilon_{u}}{\sigma_{qu}}\frac{\epsilon_{n+1}\cdot \epsilon_{j}}{\sigma_{qj}}\det\left(\Psi'^{q,u+n+1}_{q,j+n+1}\right)\\
&+\sum_{{q=1}}^{n}\frac{(\epsilon_{n+1}\cdot \epsilon_{q})^2}{(k_{n+1}\cdot k_{q})}\sum_{{i=1}\atop{i\neq 2,3,q}}^{n}\sum_{{j=1}\atop{j\neq 2,3,q}}^{n}(-1)^{i+j+\theta(q+n+1,i)+\theta(q+n+1,j)}\frac{k_{n+1}\cdot k_{i}}{\sigma_{qi}}\frac{k_{n+1}\cdot k_{j}}{\sigma_{qj}}\det\left(\Psi'^{i,q+n+1}_{j,q+n+1}\right)
\end{align*}
%
%
%

\acknowledgments

I would like to thank my adviser Anastasia Volovich for suggesting this computation to me. Furthermore, I am grateful to Anastasia Volovich and Burkhard U. W. Schwab for helpful discussions. This work is supported by the US Department of Energy under contract DE-FG02-11ER41742.

\paragraph{Note}$~$\newline
When this work was being prepared for submission, I learned that C. Kalousios and F. Rojas were working on the same problem and are about to publish their results as in \cite{chris}.

%
%



\end{document}